\documentclass[journal]{IEEEtran}

\usepackage{booktabs}
\usepackage{graphicx}
\usepackage{amsmath}
\usepackage{amssymb}
\usepackage{subcaption}
\usepackage[draft]{todonotes}
\usepackage{subcaption}
\usepackage{hyperref}

\usepackage{float}

\newcommand{\etal}{\textit{et al. }}

\begin{document}

%
\title{AeroRIT: A New Scene for Hyperspectral Image Analysis}
%
%
%

\author{Aneesh Rangnekar,~\IEEEmembership{Student Member,~IEEE,}
        Nilay Mokashi,
        Emmett Ientilucci,~\IEEEmembership{Senior Member,~IEEE,}
        Christopher Kanan,~\IEEEmembership{Senior Member,~IEEE,}
        and~Matthew J. Hoffman
\thanks{This work was supported by the Dynamic Data Driven Applications Systems Program, Air Force Office of Scientific Research under Grant FA9550-19-1-0021. We gratefully acknowledge the support of NVIDIA Corporation with the donations of the Titan X and Titan Xp Pascal GPUs used for this research. A. Rangnekar, C. Kanan, and E. Ientilucci are with the Chester F. Carlson Center for Imaging Science
, Rochester Institute of Technology. Matthew J. Hoffman is with the School of Mathematical Sciences, Rochester Institute of Technology and affiliated with Chester F. Carlson Center for Imaging Science. Nilay Mokashi is an Imaging Scientist at ON Semiconductor. 
}
\thanks{Corresponding author e-mail: aneesh.rangnekar@mail.rit.edu}
}

\markboth{To appear in IEEE TGRS}%
{Shell \MakeLowercase{\textit{et al.}}: Bare Demo of IEEEtran.cls for Journals}
%




\maketitle

\begin{abstract}
We investigate applying convolutional neural network (CNN) architecture to facilitate aerial hyperspectral scene understanding and present a new hyperspectral dataset-AeroRIT-that is large enough for CNN training. To date the majority of hyperspectral airborne have been confined to various sub-categories of vegetation and roads and this scene introduces two new categories: buildings and cars. To the best of our knowledge, this is the first comprehensive large-scale hyperspectral scene with nearly seven million pixel annotations for identifying cars, roads, and buildings. We compare the performance of three popular architectures - SegNet, U-Net, and Res-U-Net, for scene understanding and object identification via the task of dense semantic segmentation to establish a benchmark for the scene. To further strengthen the network, we add squeeze and excitation blocks for better channel interactions and use self-supervised learning for better encoder initialization. Aerial hyperspectral image analysis has been restricted to small datasets with limited train/test splits capabilities and we believe that \textit{AeroRIT} will help advance the research in the field with a more complex object distribution to perform well on. The full dataset, with flight lines in radiance and reflectance domain, is available for download at \url{https://github.com/aneesh3108/AeroRIT}. This dataset is the first step towards developing robust algorithms for hyperspectral airborne sensing that can robustly perform advanced tasks like vehicle tracking and occlusion handling.
\end{abstract}

%
\IEEEpeerreviewmaketitle

\section{Introduction}

\IEEEPARstart{C}onvolutional neural networks (CNNs) are now being widely used for analyzing remote sensing imagery and though they have achieved some success, even the most well-designed CNNs for RGB imagery struggle to achieve a mean intersection-over-union of more than 80\% on the ISPRS aerial datasets  Vaihengen\footnote{\url{http://www2.isprs.org/commissions/comm3/wg4/2d-sem-label-vaihingen.html}} and Potsdam\footnote{\url{http://www2.isprs.org/commissions/comm3/wg4/2d-sem-label-potsdam.html}} \cite{singhBMVC18overhead, mou2019relation}. This performance is in spite of the fact that these datasets have significantly higher spatial resolution with approximate ground sampling distances of $9$cm (Vaihengen) and $5$cm (Potsdam). One potential way to develop a better classifier with is to include more discriminative signatures by moving from the RGB domain to the finer spectral resolution of hyperspectral imaging (HSI) systems. HSI systems record a contiguous spectrum, usually in steps of 1 or 5 nanometers (nm), that details the contents present in the scene and can assist in increasing discrimination capability. Despite the potential for improved spectral features, HSI data has been largely unused in machine learning applications. Partially, this is because HSI sensors are significantly more expensive than their RGB counterparts, leading to HSI data being restricted to domains such as precision agriculture and environmental monitoring. The largest popular dataset in hyperspectral remote sensing for scene understanding - University of Pavia - has a spatial resolution of only 610 $\times$ 340, out of which nearly $80\%$ samples are in the \textit{undefined} class. The lack of data has made developing and training a CNN to leverage the addition spectral features of HSI prohibitively difficult. In this paper, we seek to extend CNN-based architectures developed for medical domain \cite{ronneberger2015u} and RGB domain \cite{badrinarayanan2015segnet, isola2016image} to use the additional spectral signatures of HSI data. Training such networks requires a lot of HSI data, so we introduce AeroRIT, a dataset nearly 8 times larger than the University of Pavia, and with only $17\%$ pixels under the undefined class category. While it is possible to analyze the data on a per-pixel basis, the wide variety of object distribution calls for a more structure-aware approach and hence we adopt semantic segmentation as the task of interest for this paper.

\begin{figure}
  \centering
  \begin{subfigure}{\linewidth}
    \centering
    \includegraphics[width=.9\linewidth]{Figures/image_rgb-min.pdf}
    \caption{RGB rendered version of the scene}
    \label{fig:dataset-image}
  \end{subfigure}

  \begin{subfigure}{\linewidth}
    \centering
    \includegraphics[width=.9\linewidth]{Figures/labels-min.pdf}
    \includegraphics[width=.9\linewidth]{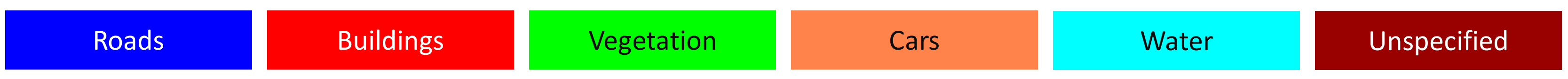}
    \caption{Semantic labeling for the scene}
    \label{fig:dataset-label}
  \end{subfigure}  
  \caption{The AeroRIT scene overlooking Rochester Institute of Technology's university campus. The spatial resolution is 1973 $\times$ 3975 pixels and covers the spectral range of 397 nm - 1003 nm in $1$ nm steps.}  
\end{figure}

There has been interest in using CNNs for analyzing remote sensing imagery \cite{uzkent2018tracking, hughes2018identifying, kleynhans2017predicting, kemker2017self, kemker2018algorithms}. Uzkent \etal adapted correlation filters trained on RGB images with HSI bands to successfully track cars in moving platform scenarios \cite{uzkent2018tracking}. Hughes \etal used a siamese CNN architecture to match high resolution optical images with their corresponding synthetic aperture radar images \cite{hughes2018identifying}. Kleynhans \etal compared the performance of predicting top-of-atmosphere thermal radiance by using forward modeling with radiosonde data and radiative transfer modeling (MODTRAN \cite{berk2014modtran}) against a  multi-layer perception (MLP) and CNN and observed better performance from MLP and CNN in all experimental cases \cite{kleynhans2017predicting}. Kemker \etal used multi-scale independent component analysis and stacked convolutional autoencoders as self-supervised learning tasks before performing semantic segmentation on multispectral and hyperspectral imagery \cite{kemker2017self, kemker2018algorithms}.

The top three hyperspectral remote sensing datasets - Indian Pines, Salinas and University of Pavia, have nearly distinctive classes and hence, learning a discriminative boundary is relatively easier without the need for advanced architectures. One of the primary reasons for lack of HSI datasets is the cost associated with its collection - the costs of hardware and flight time are very expensive, and the data collect itself is weather dependant. Assuming the costs can be offset by justifying the requirements of the task, we are also faced with high variance in spectral signatures overlooking the same scene due to factors like atmospheric scattering and cloud cover. Furthermore, HSI sensors have varying filters (or spectral response curves), which make sensor configuration necessary metadata during all operations. Finally, based on the ground sampling distance (GSD) of the scene, non-nadir RGB-trained CNN features may not provide highly discriminative information as spectral information could be lost by directly downsampling the channels using techniques such as principal component analysis (PCA) to RGB color dimension space.

To test the discriminative potential for spectral data in a more difficult setting, we flew an aircraft equipped with a hyperspectral imaging system and obtained multiple flight lines at different time stamps. We chose the flight line with the best combination of spatial and spectral quality and annotated every pixel within the flight line - we named the collect 'AeroRIT' (Fig. \ref{fig:dataset-image}). We focus on being able to distinguish between 5 classes: 1) roads, 2) buildings, 3) vegetation, 4) cars, and 5) water. This is the first dataset having challenging end-members as the signatures of some classes (buildings, cars) tend to have a large manifold and hence, generalization becomes tougher.

Mixed spectra, where multiple materials can present in single-pixel subject to GSD, are particularly challenging in remote sensing imagery due to their varying nature of the occurrence. Various spectral unmixing methods (survey: Bioucas-Dias \etal \cite{bioucas2012hyperspectral}) have been applied to separate mixed pixels but most assume the composition of all elements in the scene, referred to as end-members, are previously known. However, it is impossible to have all information about end-members when the scene is constantly changing. For example, in a moving camera setup with a push-broom sensor - each scene typically contains multiple colored cars and buildings, and applying spectral unmixing becomes difficult if the end-member signatures cannot be predetermined. We do not consider pseudo end-members for the scope of this paper and we do not tackle spectral unmixing as a problem, but address mixed pixels as sensor level noise in our tasks. This is important as the goal of this paper is to understand challenges in a moving camera setup, and although the scene is constant, we imagine the scene as one of the many time steps captured from an airborne system. 

\section{Related Work}
\subsection{Datasets for Hyperspectral Remote Sensing Imagery}
\label{sec:rw}

\begin{table*}
\tiny
\caption{Popular benchmark HSI datasets used for semantic segmentation (or pixel classification), with information on the spatial and spectral resolution. Our dataset is highlighted and as observed, is significantly bigger than its counterparts. (Acronyms: AVIRIS - Airborne visible/infrared imaging spectrometer, ROSIS - Reflective Optics System Imaging Spectrometer, HYDICE - Hyperspectral digital imagery collection experiment)}
\centering
\resizebox{0.98\textwidth}{!}{
\begin{tabular}{lccccc}
\hline
\textbf{Dataset} & Sensor & \begin{tabular}[c]{@{}c@{}}Spatial Dimensions\\ {[}px{]}\end{tabular} & \begin{tabular}[c]{@{}c@{}}Spectral Dimensions\\ {[}nm{]}\end{tabular} & \begin{tabular}[c]{@{}c@{}}Spectral \\ Bands\end{tabular} & No. of classes \\ \hline
Indian Pines & AVIRIS & 145 $\times$ 145 & 400 - 2500 & 224 & 16 \\
Salinas Valley & AVIRIS & 512 $\times$ 217 & 400 - 2500 & 224 & 16 \\
Univ. of Pavia & ROSIS & 610 $\times$ 340 & 430 - 838 & 103 & 9 \\
KSC & AVIRIS & 512 $\times$ 614 & 400 - 2500 & 224 & 13 \\
Samson & - & 952 $\times$ 952 & 401 - 889 & 156 & 3 \\
Jasper Ridge & AVIRIS & 512 $\times$ 614 & 380 - 2500 & 224 & 4 \\
Urban & HYDICE & 307 $\times$ 307 & 400 - 2500 & 210 & 6 \\ [0.1cm]
\textbf{AeroRIT} & \begin{tabular}[c]{@{}l@{}} \textbf{Headwall} \\ \textbf{Micro E}\end{tabular} & \textbf{1973 $\times$ 3975} & \textbf{397 - 1003} & \textbf{372} & \textbf{5} \\ \hline
\end{tabular}}

\label{tab:dataset}
\end{table*}

Table \ref{tab:dataset} briefly reviews the current extent of aerial hyperspectral datasets available for analysis. Other hyperspectral datasets include ICVL (Arad and Ben-Shahar \cite{arad2016sparse}) and CAVE (Yasuma \etal \cite{yasuma2010generalized}) - however, we do not include them in the table as they are non-nadir and do not have pixel-wise labels for the data. The most commonly used aerial datasets are (1) Indian Pines, (2) Salinas Valley, and (3) Univ. of Pavia. The first two primarily contain vegetation and the third contains classes typically found around a university - for example, trees, soil, and asphalt. In all three cases, the small spatial extent often leads researchers to use Monte-Carlo (MC) cross-validation splits for benchmarking the performance of various CNN-based architectures. Recently, Nalepa \etal showed that MC splits can often lead to near-perfect results as there tends to be pixel overlap (leakage) between the training and test sets \cite{nalepa2019validating}. The paper also introduces a new routine that ensures minimum to no leakage between generated data splits.  However, there is still a possibility of the network overfitting on the training set as the number of samples is significantly small (Table \ref{tab:dataset}). In our scene, we label every pixel of the flight line and create an overall hyperspectral dataset package that contains the radiance image, reflectance image, and the semantic label for every pixel. We also provide a training, validation and test set that can be used to benchmark performance without the need for cross-validation splits. We describe the data collection for our scene in Section \ref{sec:datalabel}.

\subsection{Semantic Segmentation}
\label{sec:semseg}

Semantic segmentation in HSI is often treated as a pixel classification problem due to a lack of sufficient samples. Most approaches fall under three categories: (1) spectral classifiers, (2) spatial classifiers and (3) spectral-spatial classifiers. Hu \etal used 1D-CNNs to extract the spectral features of HSIs and establish a baseline \cite{hu2015deep}. The 1D-CNN takes a pixel spectral vector as an input, followed by a convolution layer and a max pooling layer to compute a final class label. Li \etal proposed to extract pixel-pair features and treats classification as a Siamese network problem \cite{li2016hyperspectral}. Hao \etal designed a two-stream architecture, where \textit{stream1} used a stacked denoising autoencoder to encode the spectral values of each input pixel of a patch and \textit{stream2} used a CNN to process the patch's spatial features \cite{hao2017two}. Zhu \etal used a generative adversarial networks (GANs) to create robust classifiers of hyperspectral signatures \cite{zhu2018generative}. Recently, Roy \etal proposed using a 3D-CNN followed by a 2D-CNN to learn better abstract level representations for HSI scenes \cite{roy2019hybridsn}. We refer readers to Li \etal for an in-depth overview of recent methods for HSI classification \cite{li2019deep}. As the above methods do not perform semantic segmentation in the truest sense (classification: encoder $\rightarrow$ class label, segmentation: encoder $\rightarrow$ decoder), we do not include them in our network comparisons.

\section{AeroRIT}
\label{sec:datalabel}

\begin{figure}
    \centering
    \begin{minipage}[b]{0.95\linewidth}
        \begin{subfigure}[b]{.33\linewidth}
            \centering
            \includegraphics[width=.99\textwidth]{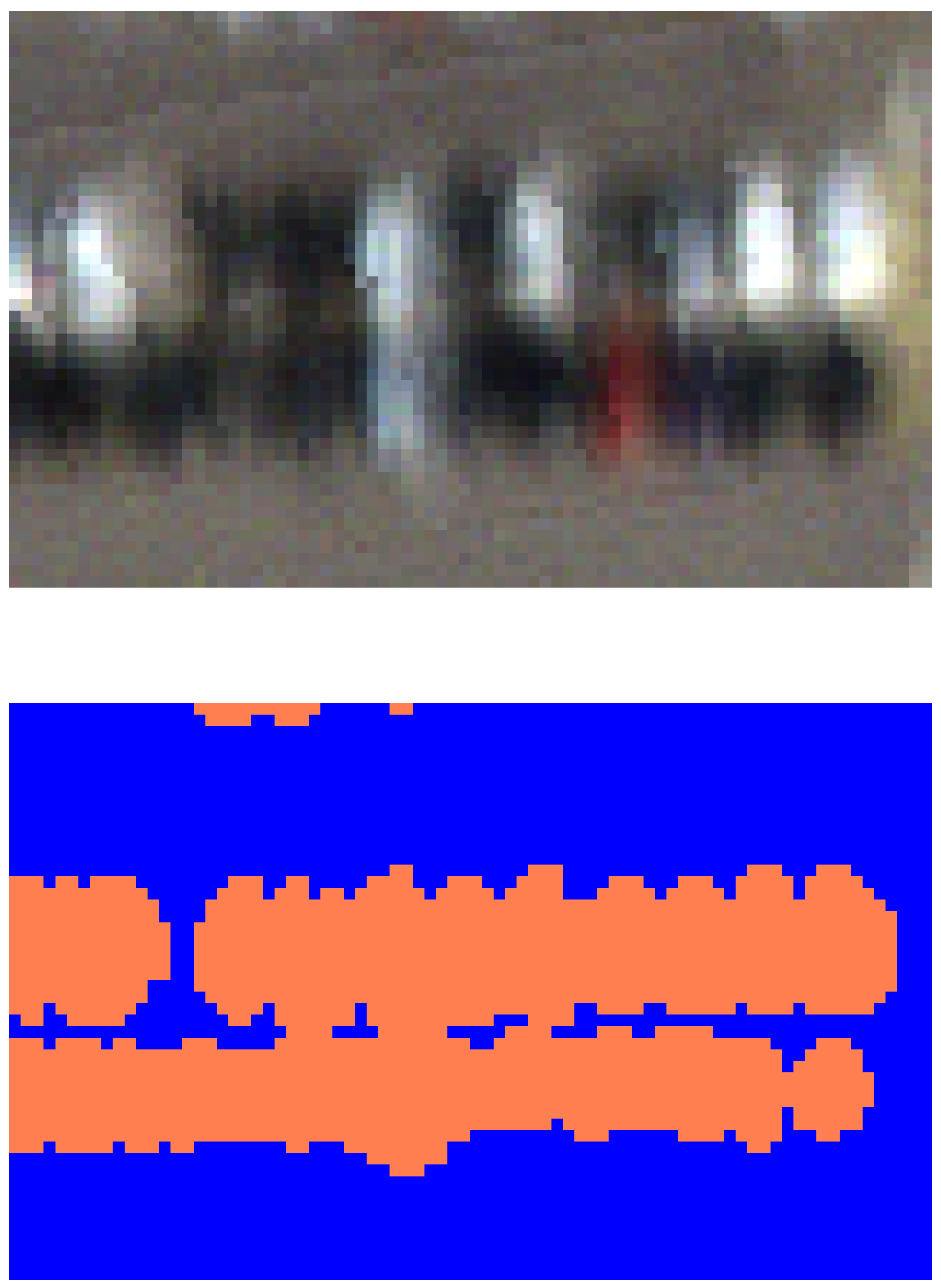}
            \caption{}
        \end{subfigure}%
        \begin{subfigure}[b]{.33\linewidth}
            \centering
            \includegraphics[width=.99\textwidth]{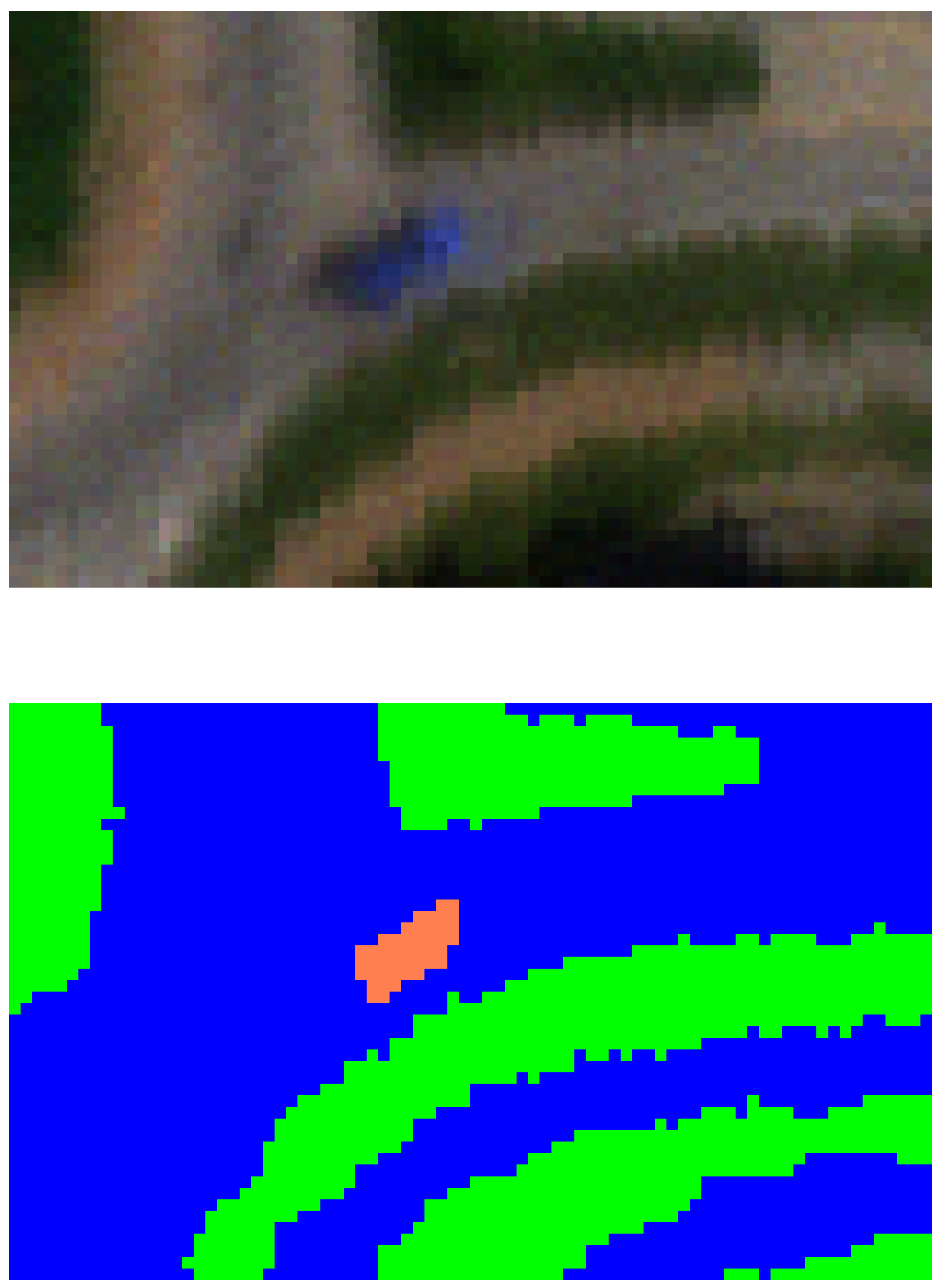}
            \caption{}
        \end{subfigure}%
        \begin{subfigure}[b]{.33\linewidth}
            \centering
            \includegraphics[width=.99\textwidth]{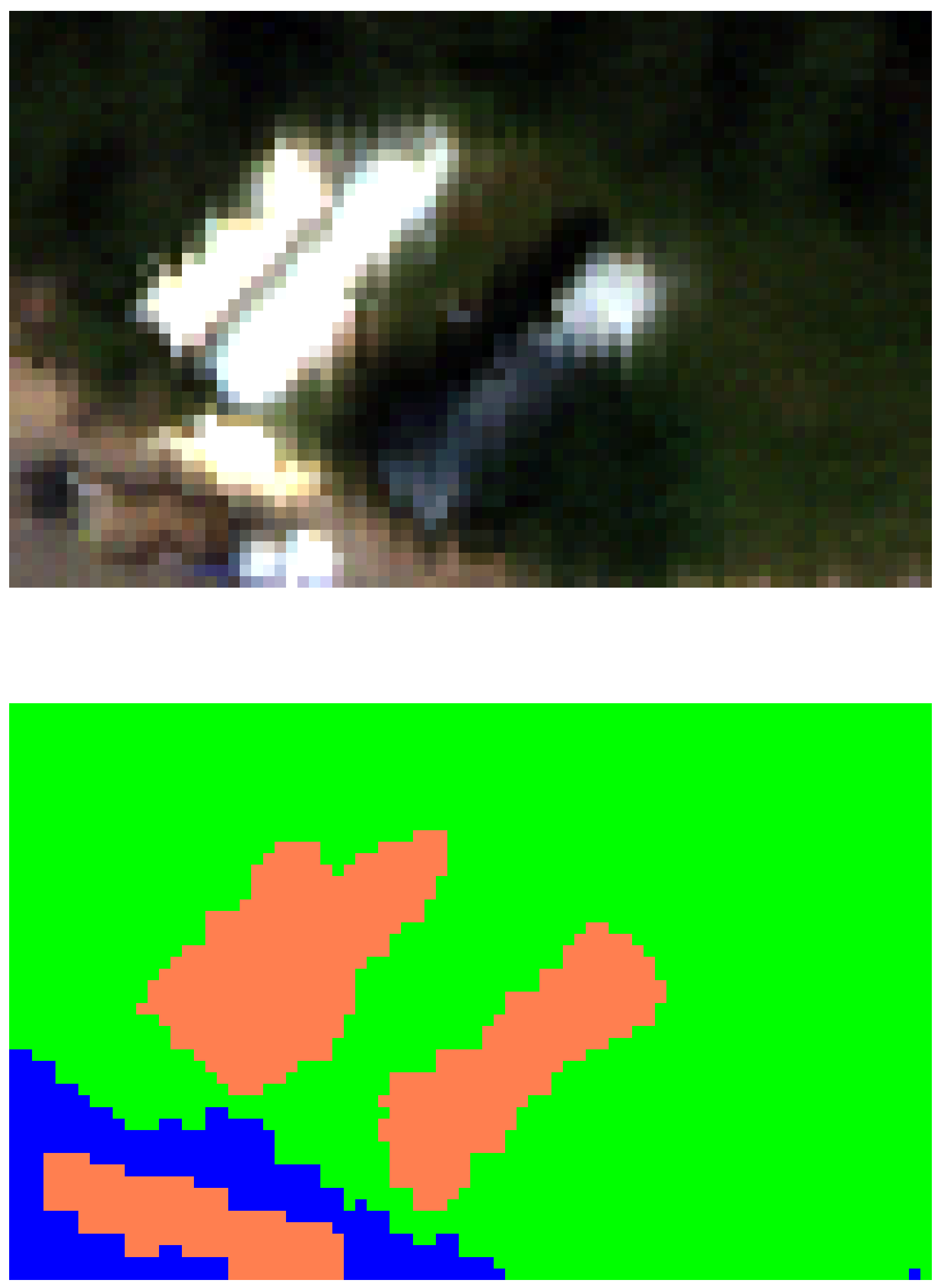}
            \caption{}
        \end{subfigure}\\
        \begin{subfigure}[b]{.33\linewidth}
            \centering
            \includegraphics[width=.99\textwidth]{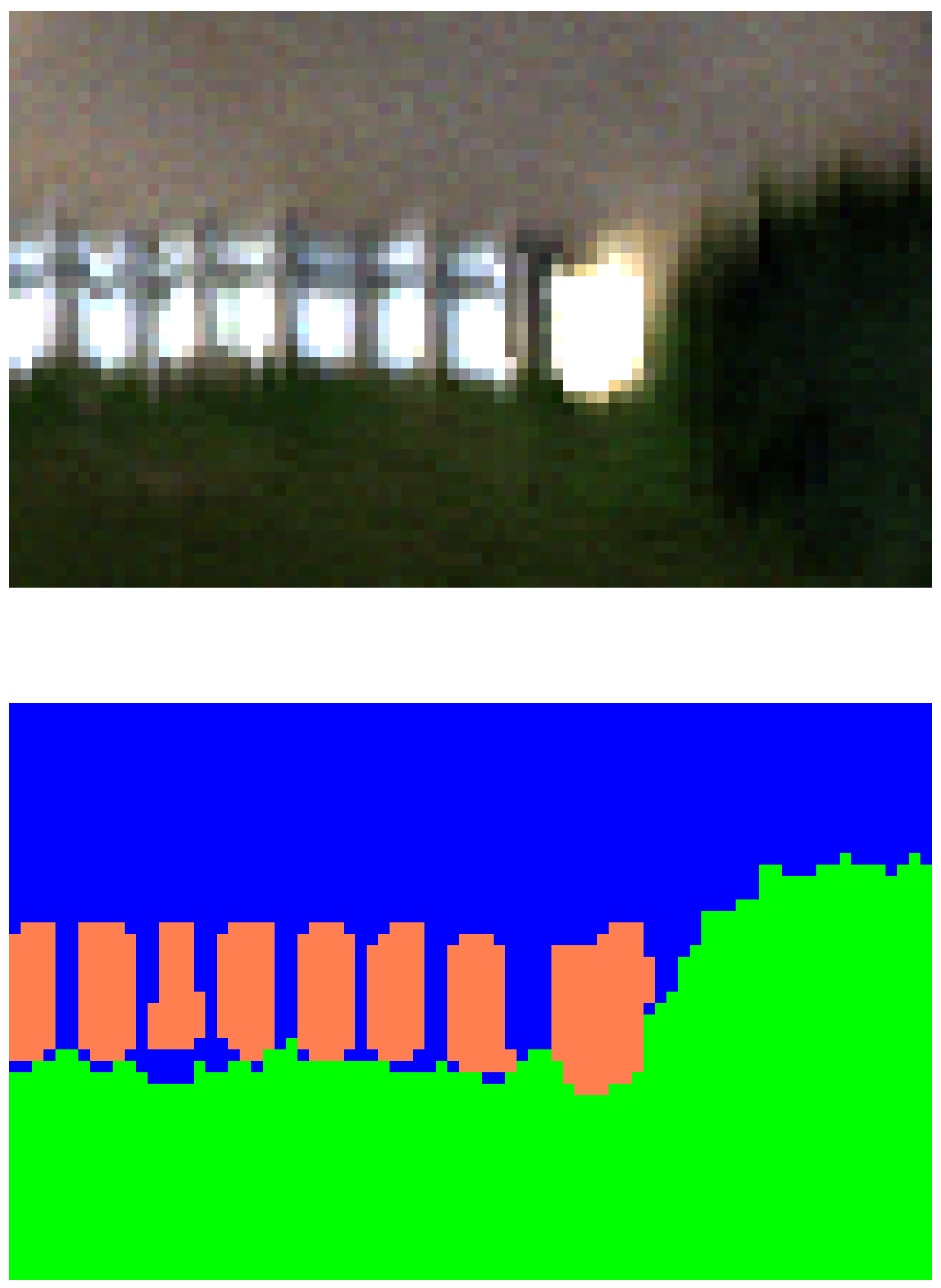}
            \caption{}
        \end{subfigure}%
        \begin{subfigure}[b]{.33\linewidth}
            \centering
            \includegraphics[width=.99\textwidth]{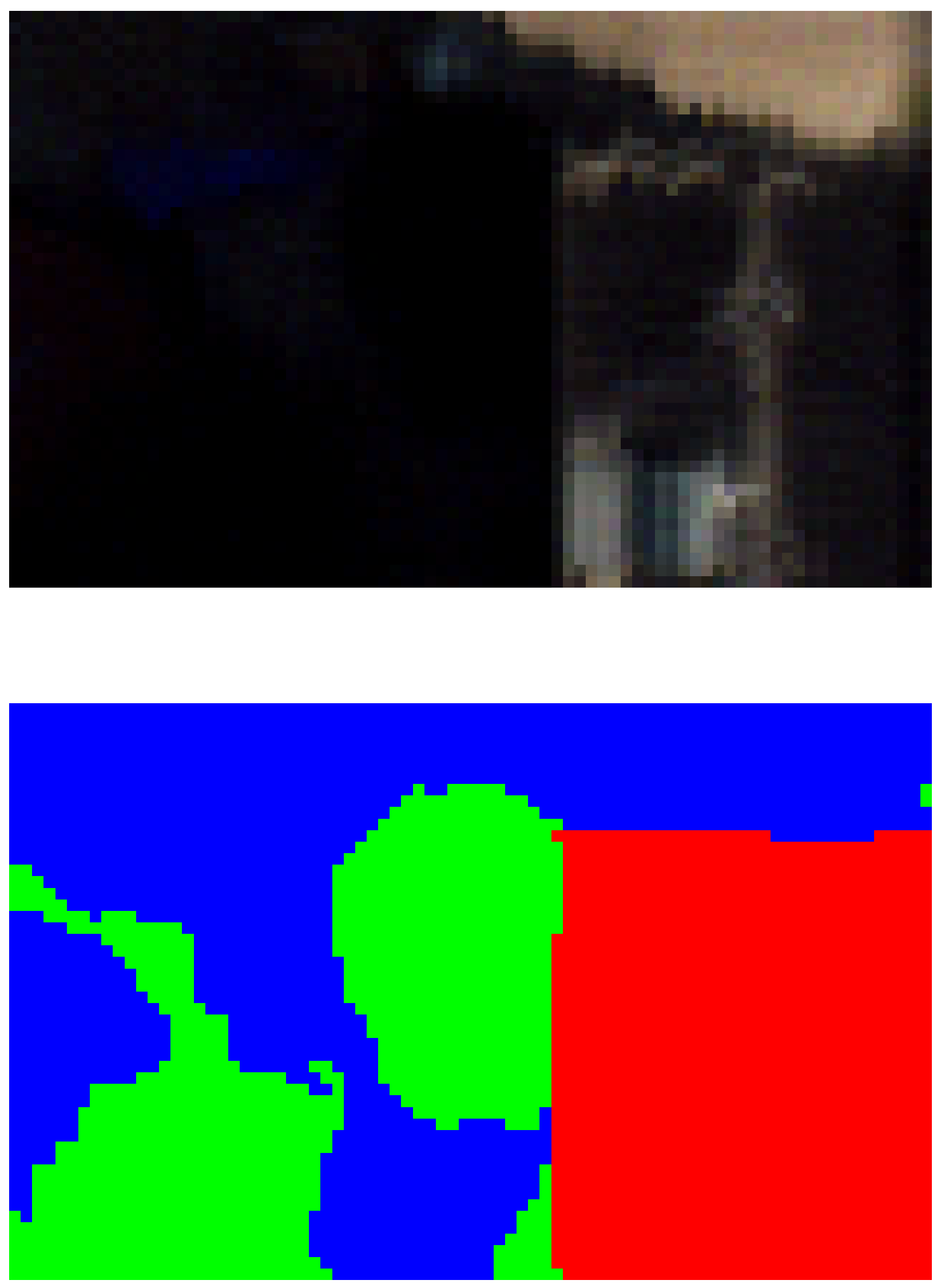}
            \caption{}
        \end{subfigure}%
        \begin{subfigure}[b]{.33\linewidth}
            \centering
            \includegraphics[width=.99\textwidth]{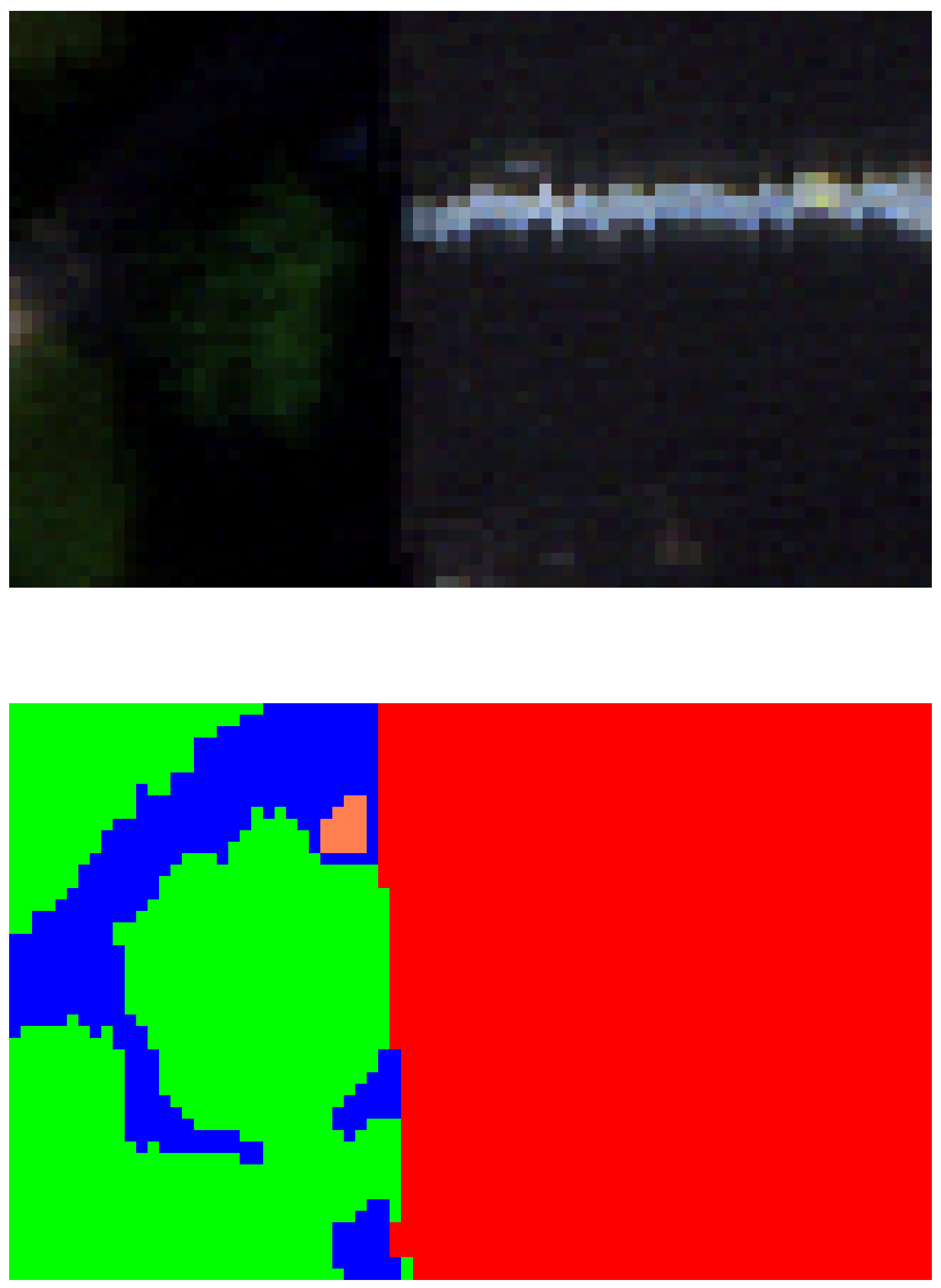}
            \caption{}
        \end{subfigure}%
    \end{minipage}%
    \caption{Challenges present in the AeroRIT scene: (a,b) low resolution, (c,d) glint, and (e,f) shadow. Each figure shows the RGB-visualized hyperspectral chip and its corresponding semantic map.}
    \label{fig:dataset-problems}
\end{figure}

The AeroRIT scene was captured by flying two types of camera systems over the Rochester Institute of Technology's university campus in a Cessna aircraft. The first camera system consisted of an 80 megapixel (MP), RGB, framing-type silicon sensor while the second system consisted of a visible near-infrared (VNIR) hyperspectral Headwall Photonics Micro Hyperspec E-Series CMOS sensor. The entire data collection took place over a couple of hours where the sky was completely free of cloud cover, except the last few flight lines at the end of the day where there was some sparse cloud cover. The aircraft was flown over the campus at an altitude of approximately 5,000 feet, yielding an effective GSD of about 0.4m for the hyperspectral imagery. The RGB data was ortho-rectified onto the Shuttle Radar Topography Mission (SRTM) v4.1 Digital Elevation Model (DEM) while the HSI was rectified onto a flat plane at the average terrain height of the flight line (that is, a low resolution DEM). Both data sets were calibrated to spectral radiance in units of $W m^{-2} sr^{-1} \mu m^{-1}$. The pixels were labeled with ENVI\footnote{data analyses were done using ENVI version 4.8.2 (Exelis Visual Information Solutions, Boulder, Colorado).}, using individual hyperspectral signatures and the geo-registered RGB images as references. As the RGB images do not form a continuous flight line (framing camera pattern) and are more in short burst captures format, we only labeled the hyperspectral scene and use it in our analysis. 

Some important challenges associated with the scene are:
\begin{itemize}
    \item \textbf{Low-resolution:} CNNs have been known to learn edge and color related features in the early to mid layers \cite{zeiler2014visualizing}. In our case, the low pixel resolution coupled with mixed pixels, makes discriminative feature learning relatively difficult. (Fig. \ref{fig:dataset-problems}a, \ref{fig:dataset-problems}b)
    \item \textbf{Glint:} Sun glint occurs due to bidirectional reflectance and the surface paint directly reflecting sunlight into the camera sensor. We observe this only occurs in certain parts of the imagery and is almost always associated with vehicles. As identifying pixels of the vehicles class is one of the end objectives, handling glint is an important topic. (Fig. \ref{fig:dataset-problems}c, \ref{fig:dataset-problems}d)
    \item \textbf{Shadows:} High rise structures (trees, buildings) often cast shadows that act as natural occlusions in scene understanding. Fig. \ref{fig:dataset-problems}f shows an image where a car is stationed right beside a building, but is nearly invisible to the human eye.
\end{itemize}

\begin{figure}
	\centering
	\includegraphics[width = 0.48\textwidth]{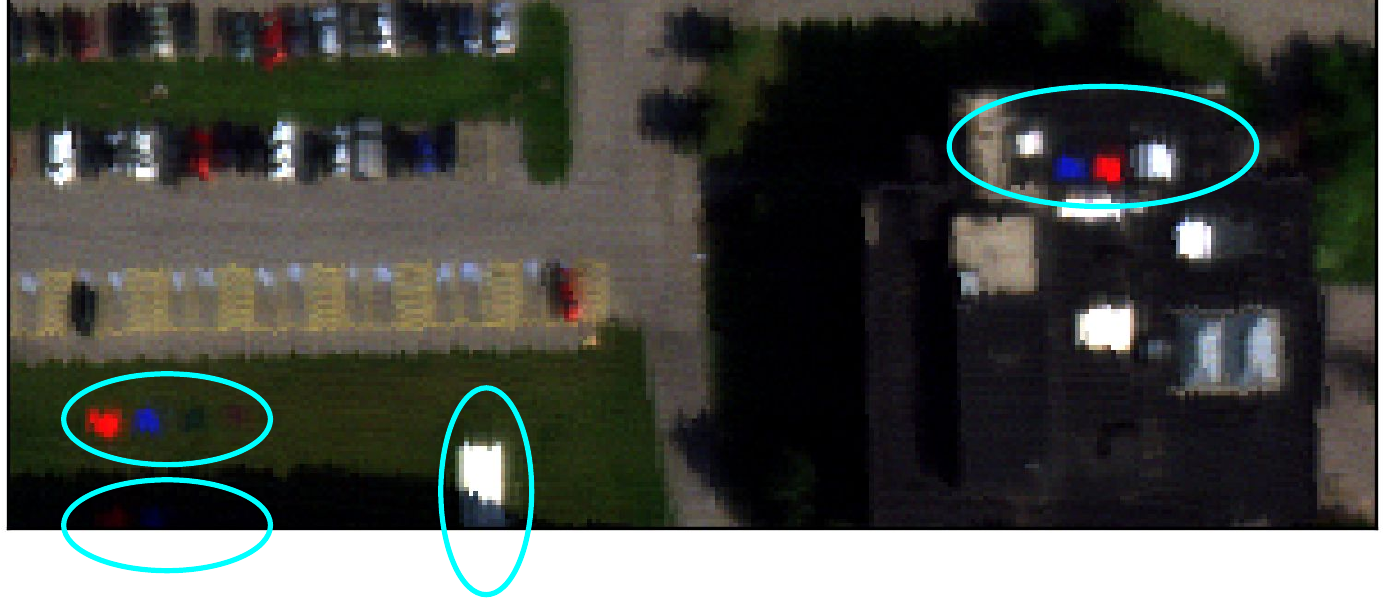}
	\caption{Targets (cyan) placed in the scene as calibration panels. We use the ground versus aerial signatures to draw a linear mapping between radiance and corresponding reflectance units.}
	\label{fig:refop}
\end{figure}

 \textbf{Conversion into reflectance data.} We calculate the surface reflectance from the calibrated radiance image using the software, ENVI. Calibration panels were deployed in the scene during the various overpasses (Fig. \ref{fig:refop}). The reflectance of these black and white uniform calibration panels was measured using a field deploy-able point spectrometer. The panels were large enough to produce full pixels in the image data (i.e., minimal pixel mixing). These full pixels enabled us to produce a linear spectral (i.e., per-band) lookup table (LUT) for the mapping of radiance to reflectance. That is, an LUT is generated for every band. This in-scene technique is often called the Empirical Line Method (ELM).  One of the key assumptions with this technique is that the atmospheric mapping of radiance to reflectance over the in-scene panels used to define the mapping, also applies, spatially, to the rest of the image.  This assumption holds fairly true for our case as the atmosphere, spatially, throughout the scene was fairly invariant and uniform.  Furthermore, the risk of multiple scattering (i.e., a non-linear issue) was very minimal due to the fact that the atmospheric conditions were so clear.
 
\begin{figure}
	\centering
	\includegraphics[width = 0.48\textwidth]{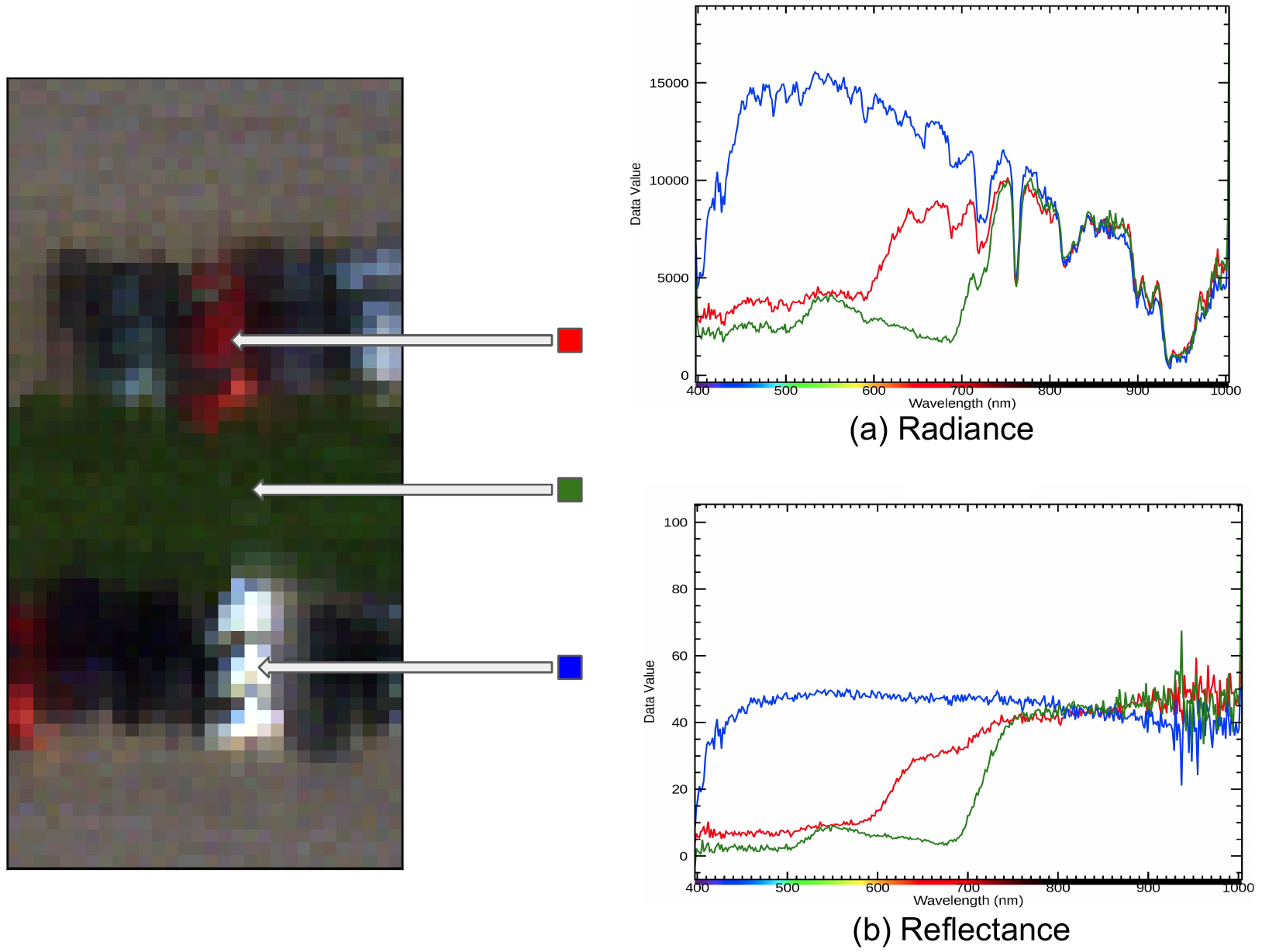}
	\caption{A comparison of signals obtained from the radiance and reflectance domains. As seen, radiance-$a$ has a varying range of amplitudes while reflectance-$b$ is restricted to the $0-100$ percentage range. The x-axis on the graph denote the bands, and the y-axis on the graph denote the value.}
	\label{fig:refrad}
\end{figure}

\section{CNN-based Network Architectures}
\label{subsec:archs}

\begin{figure}
	\centering
	\includegraphics[width = 0.45\textwidth]{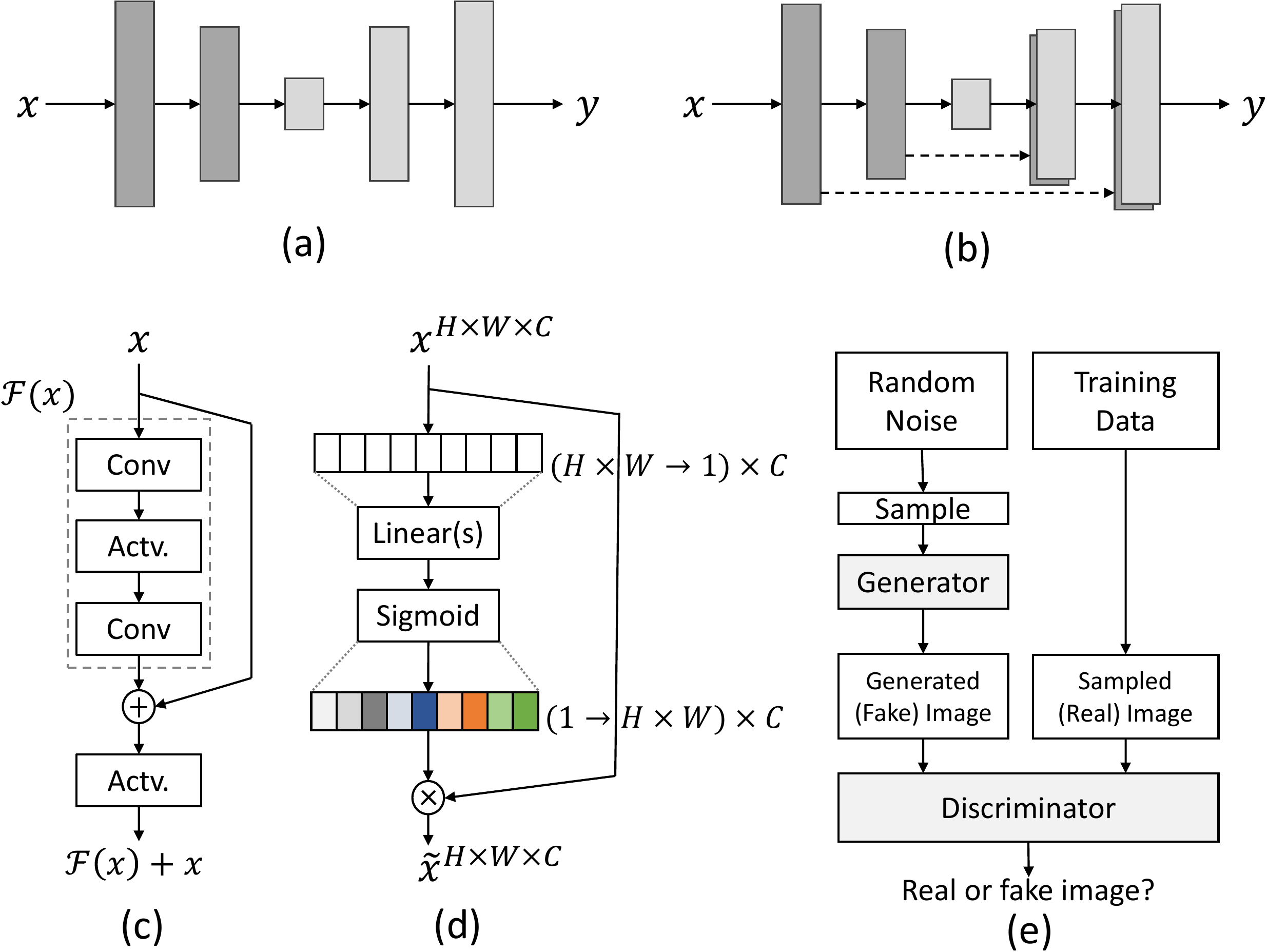}
	\caption{A graphic description of some models used in the paper. (a) SegNet \cite{badrinarayanan2015segnet}, (b) U-Net \cite{ronneberger2015u}, (c) Residual layer 	\cite{he2016deep} connection in Res-U-Net, (d) Squeeze-and-Excitation layer from SENet \cite{hu2018squeeze}, and (e) Workflow of Generative Adversarial Networks \cite{goodfellow2014generative}. Generally, the $x$ in (a), (b) is the image input, and $y$ is the mapping to be learned - e.g. depth estimation, semantic segmentation, colorization.}
	\label{fig:models}
\end{figure}

we discuss the encoder-decoder based CNN architectures that are used to establish benchmarks on the AeroRIT dataset. With respect to hyperspectral imagery, the model architectures are constrained by the following requirements: (1) They should be able to process low resolution features very well due to the nature of the data, (2) They should be able to propagate information to all layers of the network so that valuable information is not lost during sampling operations, (3) They should be able to make the most out of limited data samples and, (4) They should be as lightweight with respect to parameters as possible as the data itself is too large in size. The natural choice of selection would be a U-Net (Fig. \ref{fig:models}b), as the skip connections help propagate additional information from the encoder to the decoder. In technical terms, each skip connection concatenates all channels at layer $i$ with those at layer $n-i$, where $n$ is the total number of layers. In the following sections, we consider popular architectures: SegNet, U-Net and Res-U-Net \cite{badrinarayanan2015segnet, ronneberger2015u, he2015delving} as shown in Fig. \ref{fig:models}. As there as no pretrained models available for image processing in the hyperspectral domain, we train the networks from scratch. Furthermore, we also investigate two additional approaches that have shown to work in RGB domain: (1) Squeeze and Excitation block and (2) Generative Adversarial Networks. The former is used for improving channel inter-dependencies in the network, and the latter is used for self-supervised representation learning in some cases. We further discuss the two approaches in the subsequent subsections. 

\subsection{Squeeze and Excitation block} 
This layer block (Fig. \ref{fig:models}d) was proposed by Hu \etal to scale network responses by modeling channel-wise attention weights \cite{hu2018squeeze}. This is similar to a residual layer (Fig. \ref{fig:models}c) used in ResNets, except that the latter focuses on spatial information as compared to channel information. The workings of this layer are as follows: For any given feature block $x$, it is passed through global average pooling to obtain a channel feature vector, which embeds the distribution of channel-wise feature responses (Eqn. \ref{eq:squeeze}). This is referred to as the \textbf{squeeze} block. The vector $z$ in $\mathbb{R}^C$ (where $C$ is the number of channels) is generated by squeezing $x$ through its spatial dimensions $H \times W$, such that the $c$-th element of $\mathbf{z}$ is calculated by:

\begin{equation}
    z_c = \mathbf{F}_{squeeze}(\mathbf{x}_c) = \frac{1}{H\times W}\sum_{i=1}^{H} \sum_{j=1}^{W} x_c(i,j).
    \label{eq:squeeze}
\end{equation}

This is followed by two fully connected layers ($W_1, W_2$) and a sigmoid layer ($\sigma$), in which the channel-specific weights can be learned through a self-gating mechanism (Eqn. \ref{eq:excitation}):

\begin{equation}
    \mathbf{s} = \mathbf{F}_{excite}(\mathbf{z}, W_1, W_2) = \sigma(W_2 \delta(W_1\mathbf{z})),
    \label{eq:excitation}
\end{equation}
where $\delta$ refers to the ReLU non-linearity \cite{nair2010rectified}, $W_1 \in \mathbb{R}^{\frac{C}{r} \times C}$, $W_2 \in \mathbb{R}^{C \times \frac{C}{r}}$ and $r$ is the reduction ratio to vary the capacity of the block. This is referred to as the \textbf{excitation} block. The output of the squeeze-and-excitation block is obtained by reshaping the learned channel weights (Eqn. \ref{eq:excitation}) to the original spatial resolution and multiplying with the feature block: 

\begin{equation}
    \widetilde{\mathbf{x}_c} = s_c\,\mathbf{x}_c.
    \label{eq:se-final}
\end{equation}

The final representation $\widetilde{\mathbf{x}}$ is the combination of all $\widetilde{\mathbf{x}_c}$ (Eqn. \ref{eq:se-final}) and provides a more effective channel-weighted feature map that can be passed to the next set of layers.

\subsection{Conditional Generative Adversarial Networks}
Conditional GANs (cGANs) were first proposed by Mirza and Osindero \cite{mirza2014conditional}, and have been used widely for generating realistic looking synthetic images \cite{johnson2016perceptual,zhang2016stackgan,ledig2016photo,isola2016image}. We first discuss the base generative adversarial network (GAN) and then proceed to cGANs framework. A typical GAN (Fig. \ref{fig:models}e) consists of a generator (G) and a discriminator (D), both modeled by CNNs, tasked with learning meaningful representations to surpass each other. The generator learns to generate new \textit{fake} data instances (e.g. images, audio signals) that cannot be distinguished from the \textit{real} instances, while the discriminator learns to evaluate whether each instance belongs to the actual training dataset or is fake/synthetic (created by the generator). Formally, we can write the objective loss function as: 

\begin{equation}
    \begin{split}
        \mathcal{L}_{GAN}(G,D) = \mathbb{E}_{y}[\log D(y)] \: +
        \\
        \mathbb{E}_{z}[\log (1-D(G(z))],
    \end{split}
    \label{eq:gan_maineq}
\end{equation}
where the input to $G$ is sampled from a noise distribution $z$ (e.g. normal, uniform, spherical) and $\mathbb{E}_{y}$ is the expectation over the sample distribution (in this case, $y$). The generator learns a mapping $G: z\rightarrow y$ and tries to minimize the loss, while the discriminator tries to maximize it. 

In a cGAN setting, the input to generator is no longer just from a noise distribution, but instead appended with a source label $x$. It now learns a mapping $G: \{x,z\}\rightarrow y$ and the corresponding loss function becomes: 
\begin{equation}
    \begin{split}
        \mathcal{L}_{cGAN}(G,D) = \mathbb{E}_{x,y}[\log D(x,y)] \: + 
        \\
        \mathbb{E}_{x,z}[\log (1-D(x,G(x,z))].
    \end{split}
    \label{eq:cgan_maineq}
\end{equation}
Eqn. \ref{eq:cgan_maineq} shows us that the source label $x$ is also passed on the discriminator, which uses this additional information to perform the same task as in GAN. We use the cGAN-based image to image translation framework of Isola \etal \cite{isola2016image}, with the final objective of the generator as follows: 
\begin{equation}
    \ G^* = \arg \min_{G} \max_{D} \mathcal{L}_{cGAN} (G,D) + \lambda \mathcal{L}_{other}(G).
    \label{eq:loss_cGAN_generator}
\end{equation}
that is, generate samples of a quality that lowers the discriminator's ability to identify if the sample is from the real or fake distribution. The other loss in Eqn. \ref{eq:loss_cGAN_generator} is an additional term imposed on the generator, which forces the generated image to be as close to the ground truth as possible. We use the standard L1-loss as $\mathcal{L}_{other}(G)$.

\begin{figure}
	\centering
	\includegraphics[width = 0.48\textwidth]{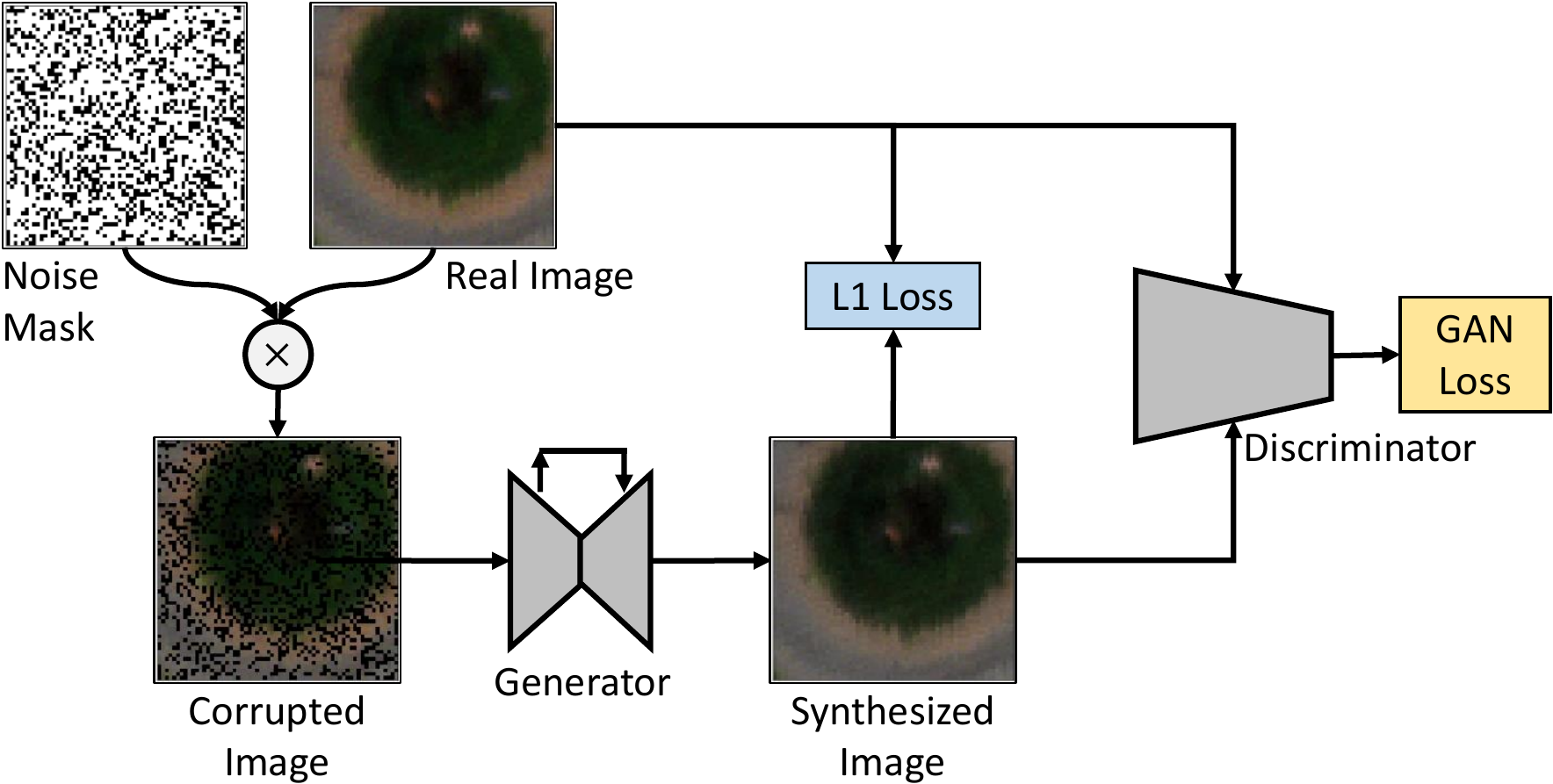}
	\caption{Procedure for image reconstruction from a corrupted image. The generator is the network under consideration (U-Net-m-SE-PReLU), and the discriminator has 5 convolution layers followed by batch normalization and leaky ReLU.}
	\label{fig:ganproc}
\end{figure}

Self-supervised learning (survey: Jing and Tian \cite{jing2019self}) has shown much potential in helping randomly initialized neural networks learn better initialization points before being applied for their original task in other domains. We apply image in-painting and image denoising as two tasks for self-supervision on our dataset. The two tasks can be described as: (1) in-painting: randomly generate binary masks and multiply them with the real image, (2) image denoising: perturb the original image with Gaussian or salt-and-pepper noise. The networks are then tasked with restoring the original image from the corrupted image. Obtaining a good quality representation of the underlying pixels in turn helps the network learn a weak prior over the image space. We adopt the above discussed cGAN framework and experiment with both the tasks. The entire training framework is summarized in Fig. \ref{fig:ganproc}.

\section{Experiments and Results}

\subsection{Experiment Configurations}
\label{subsec:expconfigurations}

\begin{figure}[t]
	\centering
	\includegraphics[width = 0.45\textwidth]{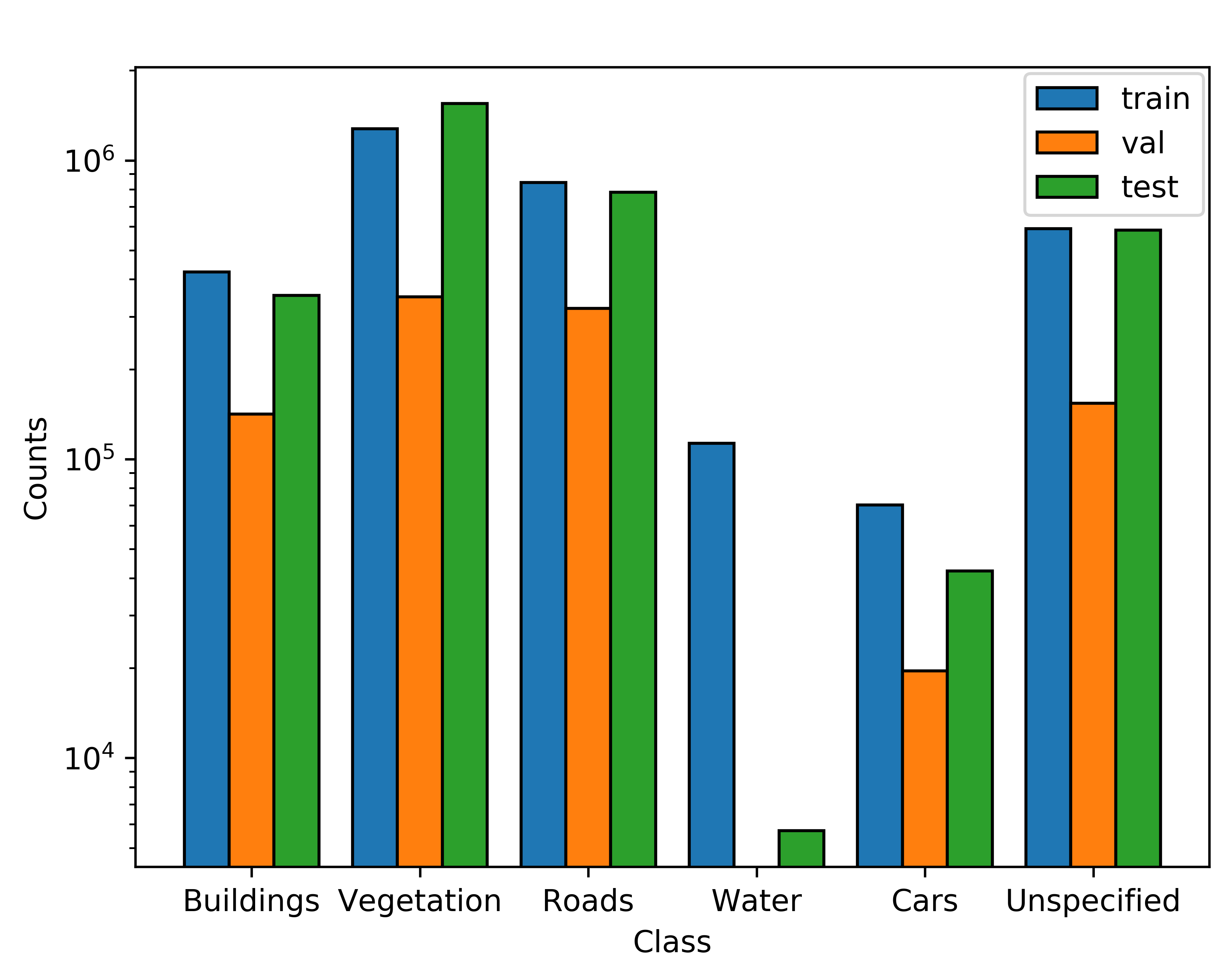}
	\caption{Label distribution of the dataset in log space. Cars are comparatively under-represented in the scene, while vegetation and roads have the highest number of samples.}
	\label{fig:dataset-stats}
\end{figure}

We use the PyTorch library \cite{paszke2017automatic} for all our experiments. We split the scene into training, validation, and test as follows: the original flight line was 1973 $\times$ 3975. We drop the first 53 rows and 7 columns and get a flight line of 1920 $\times$ 3968. We use the first 1728 columns (and all rows) for training, the next 512 columns as validation and the last 1728 as the test split. We sample 64 $\times$ 64 patches (with 50\% overlap) to create a training set and non-overlapping patches for validation and test set. Fig. \ref{fig:dataset-stats} shows the number of samples present in each class -- the scene is heavily imbalanced with reference to class \textit{cars}. We adopt basic data augmentation techniques, random flip, and rotation, and extend the dataset by a factor of four. We use a batch size of 100, and train for 60 epochs with a learning rate of $1\text{e-}4$. We also use a multi-step decay of factor $0.1$ at epoch 40 and 50.

We sample every 10th band from 400 nm to 900 nm (i.e., 400 nm, 410 nm, ..., 900 nm) to obtain 51 bands from the entire band range. As the 372 band centers are not aligned in perfect order, we use ENVI for extracting near accurate bands centers. In preliminary experiments, we found that the last set of bands (from 900 nm to 1000 nm) did not provide useful discriminative information (intuitively due to the low signal to noise ratio in the channels), so they were removed for all experiments. We normalize all data between 0 to 1 by clipping to a max value of $2^{14}$ $(16384)$.

\subsection{Loss and Metrics}
\label{subsec:loss&met}

We use weighted categorical cross-entropy to minimize the segmentation map and ignore the \textit{unspecified} class label. The weights are calculated using median frequency balancing (Eigen and Fergus \cite{eigen2015predicting}), where the number of pixels in the scene belonging into a particular class are also taken into consideration. This helps overcome the class imbalance shown in Fig. \ref{fig:dataset-stats}.

We use the following sets of metrics: overall accuracy (OA), mean per-class accuracy (MPCA), mean Jaccard Index (popularly known as mIOU) and mean Sørensen–Dice coefficient (mDICE). 




OA and MPCA report the percentage of pixels correctly classified. However, they are still slightly prone to a dataset bias when class representation is small and hence, we also report mIOU and mDICE. mIOU is the class-wise mean of the area of intersection between the predicted segmentation and the ground truth divided by the area of union between the predicted segmentation and the ground truth. Correlated to mIOU, mDICE also focuses on intersection over union and is often used as a secondary metric for measuring a network's performance on the task of semantic segmentation. We adopt mIOU as the primary metric for measure of performance. 

\subsection{Model Hyperparameters}
SegNet and U-Net both have encoders with 4 max pooling layers and gradually increasing channels by power of 2 ($C64-MP-C128-MP-C256-MP-C512-MP-$BottleNeck, where $C$ is the number of channels and $MP$ indicates a max-pooling operation). Res-U-Net blocks are built upon U-Net and, conventionally, Res-U-Net ($N$) contains $N$ identity mapping residual blocks for better information passing. In our experiments, we use $N=6$ and $N=9$ following \cite{isola2016image}. We also use smaller versions of SegNet and U-Net, called \textit{SegNet-m, U-Net-m}, that drop the number of max pooling layers from 4 to 2 to compensate for the scene's low spatial resolution and increase the channel by a factor of 2.

\subsection{Results}
\label{subsec:numbers}

We compare all the models trained for the task of semantic segmentation in Table \ref{tab:main-results}. We observe that 6-block Res-U-Net achieves the best performance, but as U-Net-m has nearly four times fewer parameters and roughly the same performance, we adopt U-Net-m as the baseline in this study. We develop on this baseline and achieve a better performing U-Net-m version that outperforms all previous baselines.

\begin{table}[t]
\centering
\caption{Performance of various models used for establishing baseline on the AeroRIT test set.}
\resizebox{0.48\textwidth}{!}{%
\begin{tabular}{@{}lcccc@{}}
\toprule
 & \begin{tabular}[c]{@{}c@{}}pixel acc. \\ (OA)\end{tabular} & \begin{tabular}[c]{@{}c@{}}mean pixel acc. \\ (MPCA)\end{tabular} & mIOU & mDICE \\ \midrule
SegNet & 92.12 & 72.97 & 52.60 & 61.50 \\
U-Net & 93.15 & 72.90 & 60.40 & 68.63 \\
Res-U-Net (6) & 93.28 & 88.09 & 72.55 & 82.15 \\
Res-U-Net (9) & 93.25 & 84.64 & 70.88 & 80.88 \\
SegNet-m & 93.20 & 74.86 & 59.08 & 67.41 \\
U-Net-m & 93.25 & 89.66 & 70.62 & 80.86 \\
\textbf{U-Net-m (ours)} & \textbf{93.61} & \textbf{90.67} & \textbf{76.40} & \textbf{85.60} \\ \bottomrule
\end{tabular}%
}
\label{tab:main-results}
\end{table}

\begin{table}[t]
\centering
\caption{Impact of each component added to the baseline U-Net-m model from Table \ref{tab:main-results}.}
\resizebox{0.48\textwidth}{!}{%
\begin{tabular}{@{}cccccc@{}}
\toprule
\begin{tabular}[c]{@{}c@{}}SE \\ layer\end{tabular} & \begin{tabular}[c]{@{}c@{}}SE act. \\ PReLU\end{tabular} & cGAN & \begin{tabular}[c]{@{}c@{}}mean pixel acc. \\ (MPCA)\end{tabular} & mIOU & mDICE \\ \midrule
 &  &  & 89.66 & 70.62 & 80.86 \\
\checkmark &  &  & 88.59 & 75.35 & 84.05 \\
\checkmark & \checkmark &  & 90.28 & 75.89 & 84.81 \\
\checkmark & \checkmark & \checkmark & \textbf{90.67} & \textbf{76.40} & \textbf{85.60} \\ \bottomrule
\end{tabular}%
}
\label{tab:netadditions}
\end{table}

\begin{figure}
	\centering
	\includegraphics[width = 0.48\textwidth]{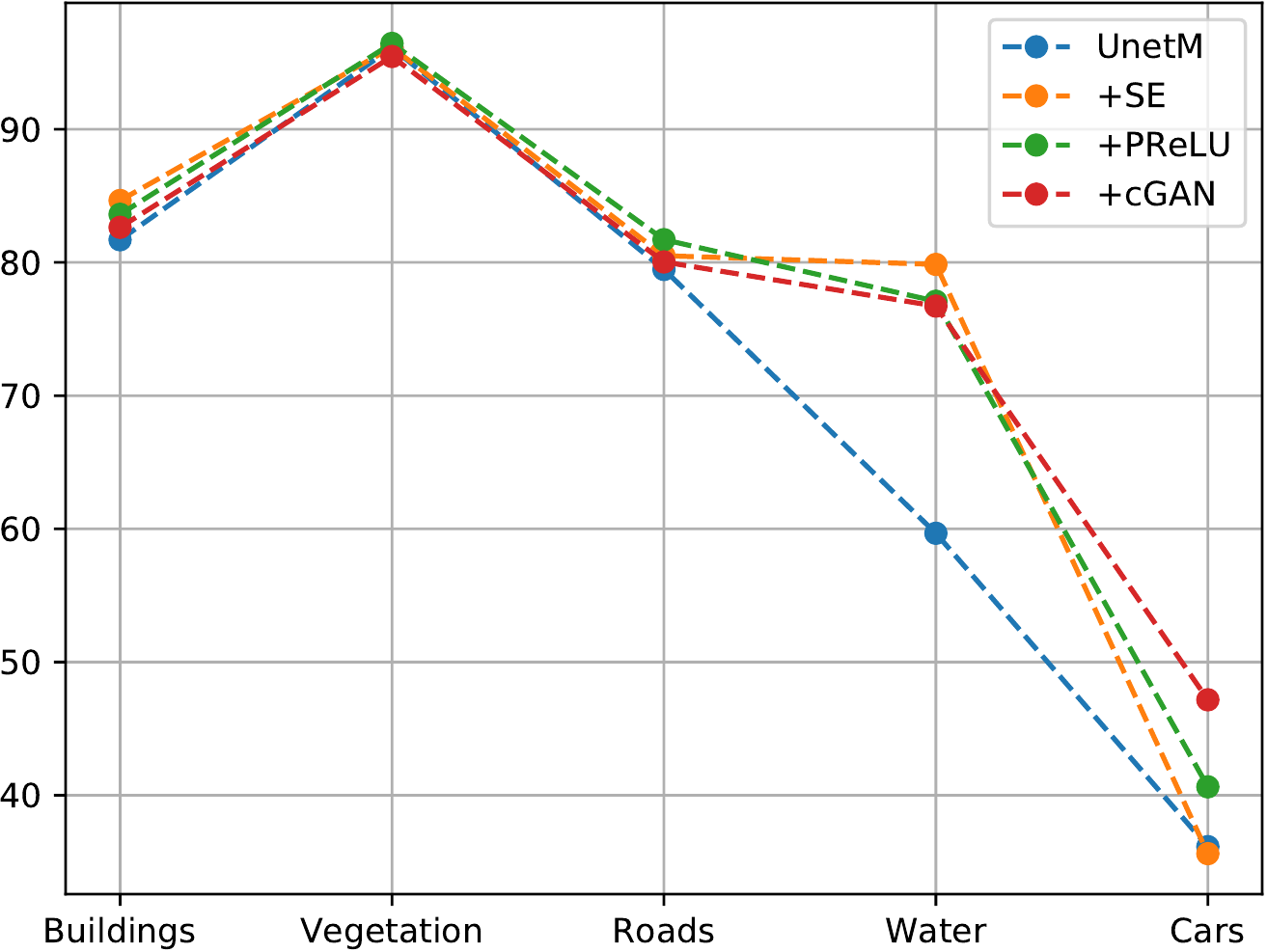}
	\caption{Results with various additions to normal U-Net-m. The y-axis is the IOU measure. SE-block and its additions improves the performance of water pixel identification by nearly 20 points over the baseline, and the overall modifications improve the performance of car pixel identification by ∼8  points. Self-supervised learning is the factor that contributes to the large improvement in car pixel identification.}
	\label{fig:netadditionstats}
\end{figure}

We discuss the approaches used to further improve the performance of the U-Net-m architecture (Table \ref{tab:netadditions}, Fig. \ref{fig:netadditionstats}). We adopt the Inception-variant of Squeeze-and-Excitation (SE) block with a reduction ratio $r = 2$ - we do not sum the output of the SE block with the original channel space as a skip connection, but use it as the importance-weighted channel output  \cite{hu2018squeeze}. We add a SE block after every $conv-batchnorm-relu$ combination on the encoder side of the network. This increases the U-Net-m performance by almost 4 points. We further replace every ReLU activation with parameterized ReLU (PReLU) and observe a slight performance boost \cite{he2015delving}. 

\begin{figure}
	\centering
	\includegraphics[width = 0.48\textwidth]{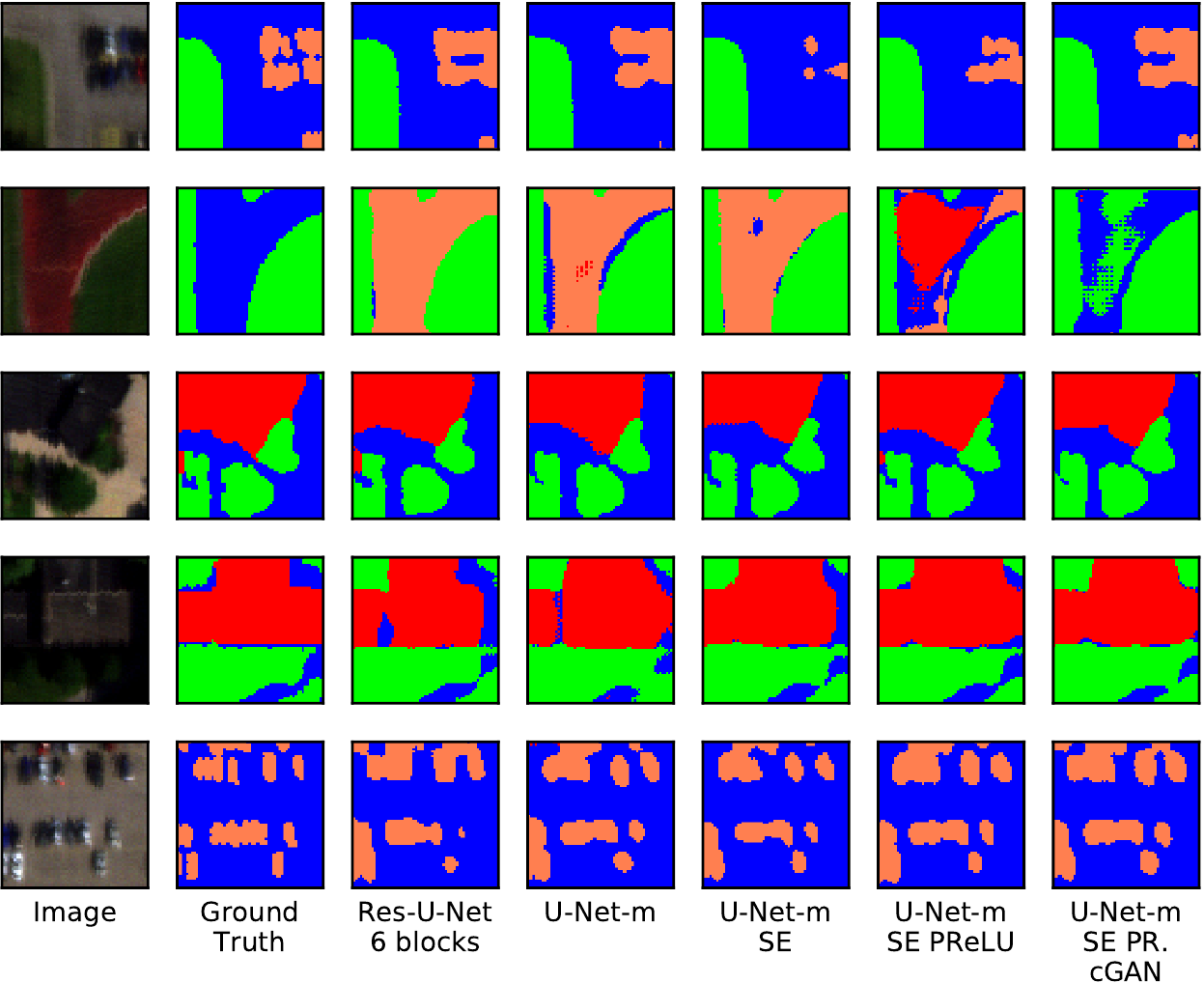}
	\caption{Successful cases: Outputs for a set of images among all networks. The racetrack image (row 2) shows that the cGAN trained network is the only one that is able to understand that the red unseen track patch is not a car or building.}
	\label{fig:goodimgs}
\end{figure}

We apply image in-painting and image denoising as the two tasks for self-supervision and we found in-painting to work as a better technique in the preliminary experiments. Once the network is trained on in-painting, we then retain the encoder weights and train the decoder from scratch on semantic segmentation. This approach further increases the performance by 1 point and has the cleanest inference labels (Fig. \ref{fig:goodimgs}).

\begin{figure}
	\centering
	\includegraphics[width = 0.48\textwidth]{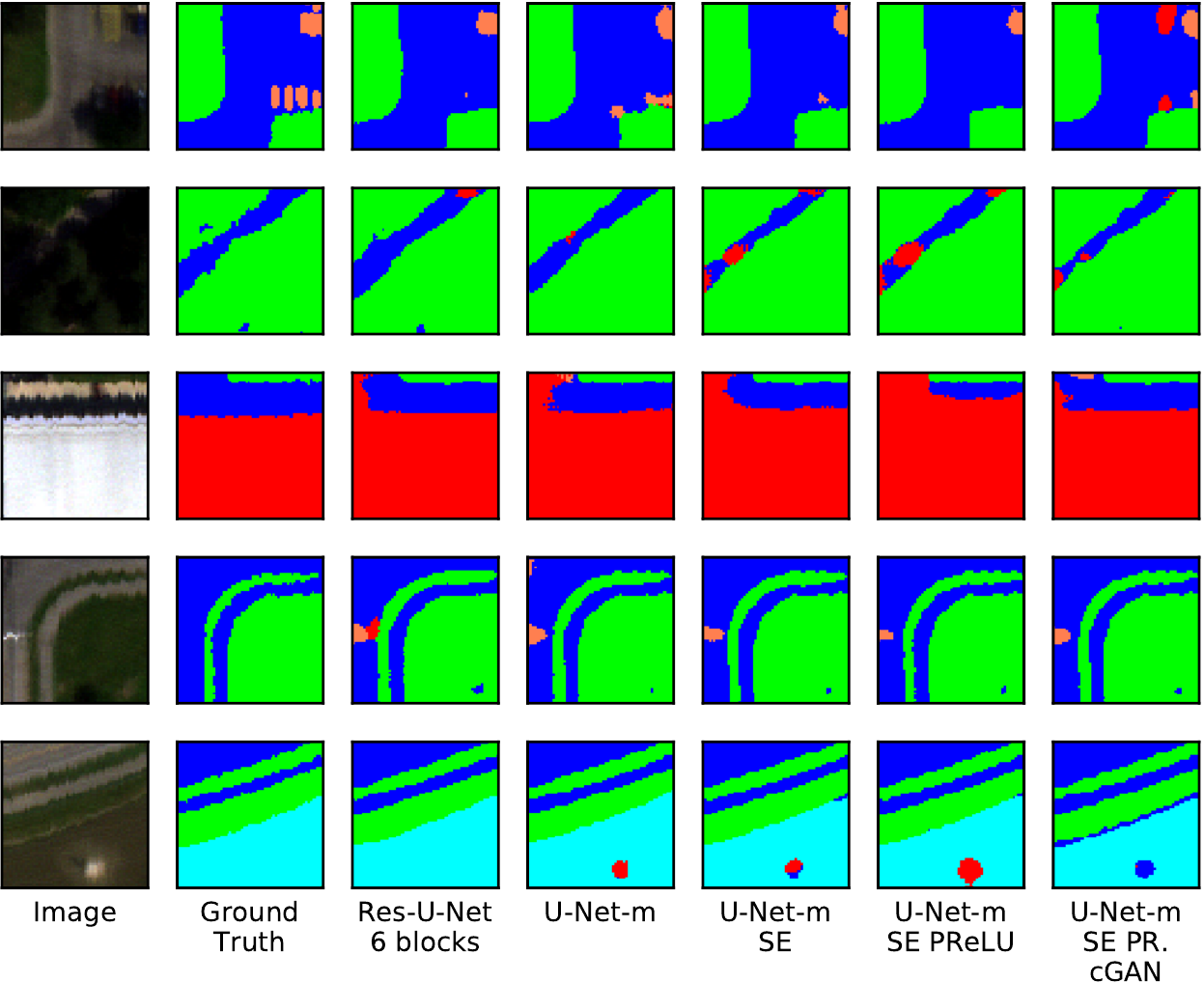}
	\caption{Failure cases: Outputs for a set of images through all trained networks. All networks predict building in between the road in Row 2, and misclassify zebra crossing as a car.}
	\label{fig:badimgs}
\end{figure}

\section{Discussions}
Along with the promising performance that CNNs have achieved in the hyperspectral domain \cite{kemker2018low,hao2017two,zhu2018generative,roy2019hybridsn}, there are several directions of research that can be construed as open problems on this dataset:

\subsubsection*{1) Radiance or Reflectance}
We compared the performance of U-Net-m-SE on the radiance and reflectance sets of images and obtain an mIOU of nearly 5 points less when using reflectance (reflectance-mIOU: $69.90$ vs radiance-mIOU: $75.35$). We hypothesize that the difference in discriminative signatures (Fig. \ref{fig:refrad}) might be one of the reasons behind the performance loss. As reflectance is the atmosphere rectified version and is theoretically less noisy, the performance drop is intriguing. However, as our primary domain of interest is the radiance space, we do not investigate this further and leave analyzing reflectance as an open problem. 

\subsubsection*{2) Data augmentation}
We identify some of the tricky cases within the flight line in Fig. \ref{fig:badimgs}. Rows $1-3$ show images partly under the shadow that have misclassified pixels, Row $4$ shows a pedestrian crossing misclassified as a vehicle and Row $5$ shows an object shining inside the fountain due to glint. The signatures vary heavily in amplitude due to shadows and glint and hence cause the networks to confuse between different classes. Conventional RGB augmentations (brightness, contrast) rely on the assumption of uniform scene illumination irrespective of image size. However, as hyperspectral signatures can vary drastically under varying atmospheric conditions and are subject to the adjacency effect, it is not possible to directly apply RGB-based augmentations for improving performance.

\subsubsection*{3) Understanding the workings of CNNs}
Learning task-relevant representations has been the forte of CNNs and a lot of techniques have been proposed to understand their internal workings \cite{zhou2016learning, selvaraju2017grad, srinivas2019fullgradient}. However, none of these techniques have been applied towards hyperspectral imagery and CNN architectures are often treated as black-box approximators giving constant performance improvements. This approach is not favorable - knowing why a particular pixel has been classified as belonging to 'cars' or why limiting max pooling layers to 2 (and indirectly, the receptive field) boosted the performance, can in turn help design better architectures. Hence, there is a need to understand the why and how of CNNs with respect to HSI: more importantly, to analyze how much of the pixels' classification depends on its spatial information (that is, structure) as compared to the spectral information (spectrum).

\section{Conclusion}

This paper introduces AeroRIT, the first large-scale aerial hyperspectral scene with pixel-wise annotations. Our scene is nearly eight times bigger than the previously largest scene and is composed of challenging factors like shadows, glint and mixed pixels that make inference difficult. We trained networks for semantic segmentation and established a baseline using squeeze and excitation block, self-supervision and PReLU activation. We believe AeroRIT can be used for future work in multiple areas of remote sensing, including but not limited to data augmentation and network architecture designing.

\appendices



\ifCLASSOPTIONcaptionsoff
  \newpage
\fi



%
\bibliographystyle{IEEEtran}
\bibliography{ref}

%






\end{document}